\documentclass[aip, amsmath,amssymb,reprint]{revtex4-1}

\usepackage{graphicx}
\usepackage{dcolumn}
\usepackage{bm}

\usepackage[utf8]{inputenc}
\usepackage[T1]{fontenc}
\usepackage{mathptmx}

\begin{document}

\preprint{AIP/123-QED}

\title{Phase equilibrium of liquid water and hexagonal ice from enhanced sampling molecular dynamics simulations}

\author{Pablo M. Piaggi}
 \affiliation{Department of Chemistry, Princeton University, Princeton, NJ 08544, USA
}
 \email{ppiaggi@princeton.edu.}
\author{Roberto Car}%
 \affiliation{Department of Chemistry and Department of Physics, Princeton University, Princeton, NJ 08544, USA}%

\date{\today}

\begin{abstract}
We study the phase equilibrium between liquid water and ice Ih modeled by the TIP4P/Ice interatomic potential using enhanced sampling molecular dynamics simulations.
Our approach is based on the calculation of ice Ih-liquid free energy differences from simulations that visit reversibly both phases.
The reversible interconversion is achieved by introducing a static bias potential as a function of an order parameter.
The order parameter was tailored to crystallize the hexagonal diamond structure of oxygen in ice Ih.
We analyze the effect of the system size on the ice Ih-liquid free energy differences and we obtain a melting temperature of 270 K in the thermodynamic limit.
This result is in agreement with estimates from thermodynamic integration (272 K) and coexistence simulations (270 K).
Since the order parameter does not include information about the coordinates of the protons, the spontaneously formed solid configurations contain proton disorder as expected for ice Ih.
\end{abstract}

\maketitle

\section{\label{sec:intro}Introduction}

The study of phase equilibria using computer simulations is of central importance to understand the behavior of a given model.
However, finding the thermodynamic condition at which two or more phases coexist is particularly hard in the presence of first order phase transitions.
In this case, the transformation between phases takes place through nucleation and growth, and this mechanism is characterized by a free energy barrier.
As a consequence of this barrier, transitions between the phases are rarely observed during a standard molecular dynamics simulation.
This lack of ergodicity renders the estimation of the free energy difference between the phases involved impossible, except in trivial models.

Here we shall focus on the case of water.
This ubiquitous and fascinating substance has at least 18 solid polymorphs that exhibit rich and diverse characteristics.
The most stable form of ice at ambient pressure is ice Ih, and it is therefore the most common polymorph in planet Earth's surface and atmosphere.
In spite of the complexity of water, the phase diagram of many water models has been carefully studied using a combination of the thermodynamic integration technique and the integration of the Clausius-Clapeyron equation \cite{Sanz04,Vega05}.
The equilibrium between liquid water and ice Ih, in particular, has also been studied using the direct coexistence technique\cite{Garcia06,Conde17} and the interface pinning method\cite{Pedersen13,Cheng19}.

In this work we employ enhanced sampling simulations to study the equilibrium between liquid water and ice Ih.
We used the TIP4P/Ice\cite{Abascal05} water model that considers rigid molecules.
In order to achieve ergodic sampling, we construct a bias potential using the variational principle of Valsson and Parrinello\cite{Valsson14}.
The bias potential is a function of an order parameter designed to distinguish between the liquid phase and ice Ih.
An important feature of this order parameter is that it drives the crystallization of ice Ih such that the crystal structure forms in an orientation compatible with the simulation box\cite{Piaggi19b}.

\section{\label{sec:crystalstructure}Crystal structure of ice I\lowercase{h}}

\begin{figure*}
\includegraphics[width=0.55\textwidth]{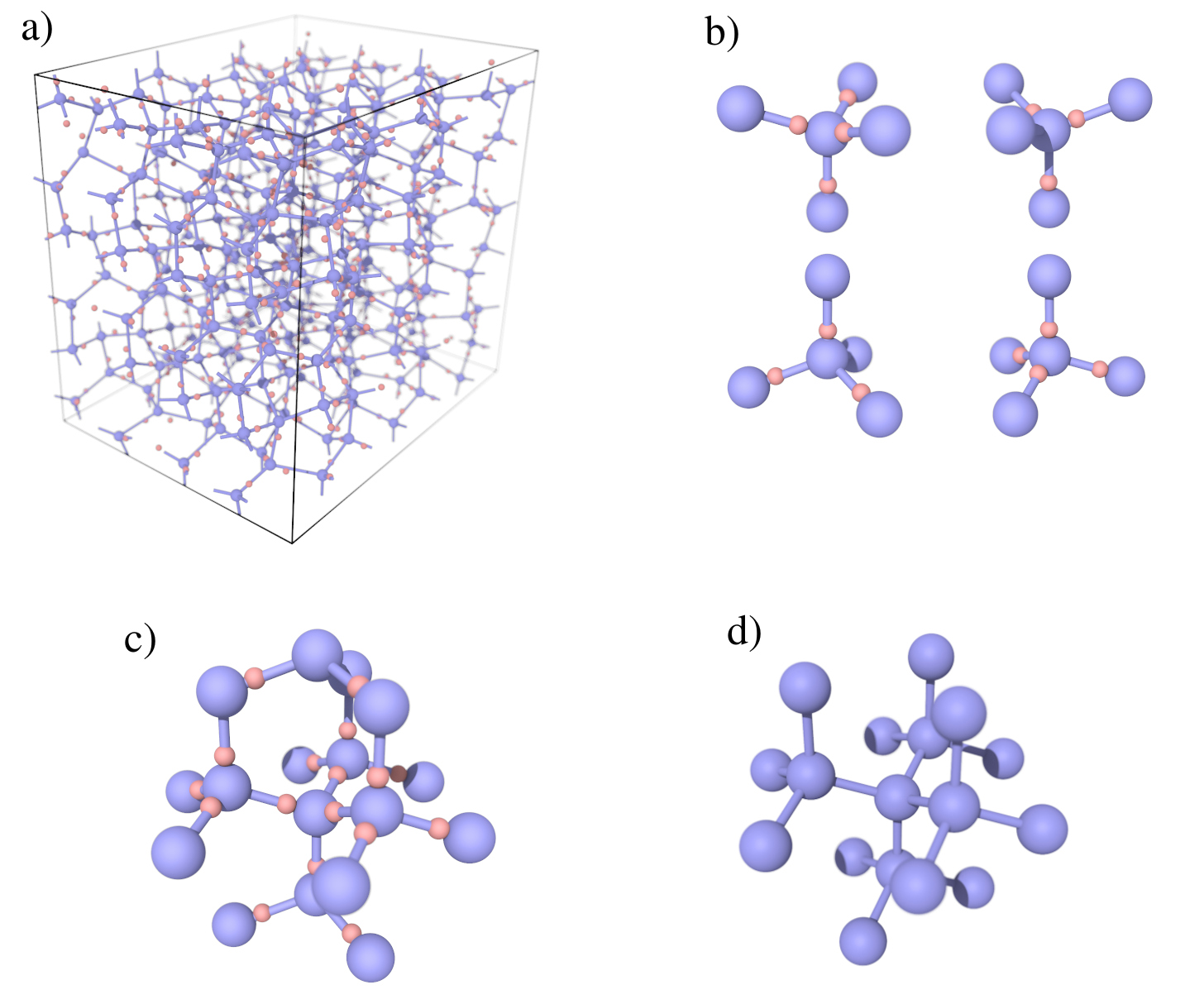}
\caption{\label{fig:Figure1} Crystal structure of ice Ih. a) 3D perspective of an ice Ih configuration with 288 water molecules at 270 K. b) The four basic environments around an oxygen atom. The proton configuration in each environment is one of the six possible choices and has been chosen randomly. c) Extended environment of ice Ih including 17 nearest neighbors. There are four equivalent environments with different orientations. d) Extended environment of ice Ic including 16 nearest neighbors. Notice the similarity with the environment of ice Ih shown in c). Images obtained with the software Ovito\cite{Stukowski09}. Oxygen atoms are shown in blue and hydrogen is shown in red. Only O-O bonds are shown for clarity.}.
\end{figure*}

Before discussing the details of the order parameter we will describe the crystal structure of ice Ih \cite{PetrenkoIce} (see Fig.~\ref{fig:Figure1}a).
Ice Ih can be thought of as a lattice of oxygen atoms arranged in the hexagonal diamond crystal structure.
In this crystal structure the basic environment around each oxygen atom has four oxygen neighbors arranged in a regular tetrahedron.
The lattice has a 4 atom basis, therefore there are 4 distinct environments if the symmetry operations of the space group are not considered.
In other words, in the crystal structure the basic environment can have 4 distinct orientations as shown in Fig.~\ref{fig:Figure1}b.
The basic environment is shared both in ice Ih and ice Ic, although these structures can be distinguished by considering extended environments (see Fig.~\ref{fig:Figure1}c and d).
 
We now consider the positions of the hydrogen atoms (protons) within the hexagonal diamond lattice of oxygen.
The positions of the protons will determine the orientation of the water molecules.
Since the identity of the water molecule is preserved in ice Ih then the protons have to satisfy the ice rules proposed by Bernal and Fowler\cite{Bernal33} and stated clearly in ref.\ \citenum{Pauling35} by Pauling. 
The ice rules state that each oxygen atom has two nearest neighbor protons, with a distance similar to the one shown in the gas or the liquid phase, and two next nearest neighbor protons. 
Alternatively one can think that each oxygen-oxygen bond in the hexagonal diamond lattice has exactly one proton that is closer to one of the two oxygen atoms.
Bernal and Fowler noted\cite{Bernal33}, however, that this prescription does not completely determine the position of the protons in the lattice.
Indeed, one of the most fascinating characteristics of ice Ih is that it contains proton disorder.
The proton disorder gives rise to an entropic contribution to the free energy called residual entropy making the entropy of ice Ih at 0 K non-zero.
Pauling\cite{Pauling35} estimated the residual entropy of ice Ih to be $k_B \log (3/2)$ per molecule and this result is often used as an input to the calculation of free energies of water using thermodynamic integration\cite{Vega05}.

It is worth noting that there is a proton ordered phase of ice in which the oxygen atoms sit in the same positions as in ice Ih\cite{Tajima82}.
This phase is called ice XI and has a net polarization along the $[0001]$ crystallographic axis.
It is thus ferroelectric.

\section{\label{sec:orderparam}On the order parameter for ice I\lowercase{h}}

In this work, we aim to perform simulations that go reversibly from the liquid to ice Ih with arbitrary proton configurations.
For this reason we will construct an order parameter that depends only on the position of the oxygen atoms.
In this way, the proton configuration will form spontaneously during the simulation without any bias towards a particular configuration.
We will employ the order parameter introduced in ref.~\citenum{Piaggi19b} and we summarize here only the most important details.

We consider the four different extended environments of the hexagonal diamond lattice $\chi_1,\chi_2,\chi_3$, and $\chi_4$.
One of these environments is shown in Fig.~\ref{fig:Figure1}c and the rest of them correspond to rotations of the one shown.
We also define a similarity kernel\cite{Bartok13} between $\chi_l \in X$ with $X = \{\chi_1,\chi_2,\chi_3,\chi_4\}$ and a generic environment $\chi$,
\begin{equation}
    k_{\chi_l}(\chi) = \int \rho_{\chi_l}(\mathbf{r}) \rho_{\chi}(\mathbf{r}) \: d\mathbf{r}
    \label{eq:kernel1}
\end{equation}
where $\rho_{\chi_l}(\mathbf{r})$ and $\rho_{\chi}(\mathbf{r})$ are the atomic densities corresponding to the environments $\chi_l$ and $\chi$, respectively.
If the densities are represented by sums of Gaussians centered at the neighbors' positions with spread $\sigma$, the kernel becomes:
\begin{equation}
    k_{\chi_l}(\chi) = \frac{1}{n} \sum_{i \in \chi_l} \sum_{j \in \chi} \exp\left(-\frac{|\mathbf{r}_i^l-\mathbf{r}_j|^2}{4 \sigma^2}\right)
    \label{eq:kernel2}
\end{equation}
where $n$ is the number of neighbors in the environment $\chi_l$, and $\mathbf{r}_i^l$ and $\mathbf{r}_j$ are the positions of the neighbors in environments $\chi_l$ and $\chi$, respectively.
In Eq.\ \eqref{eq:kernel2} we have added a normalization such that $k_{\chi_l}(\chi_l)=1$.
Now we have 4 similarity kernels that allow us to identify whether a given environment is compatible with one of the environments in ice Ih.
However, we would like a single similarity measure between a given environment and \emph{any} of the 4 environments in ice Ih.
Thus we define another kernel,
\begin{equation}
    k_X(\chi) = \max \{k_{\chi_1}(\chi),k_{\chi_2}(\chi),k_{\chi_3}(\chi),k_{\chi_4}(\chi) \}
    \label{eq:kernel_multi1}
\end{equation}
that achieves that purpose.

\begin{figure}
\includegraphics{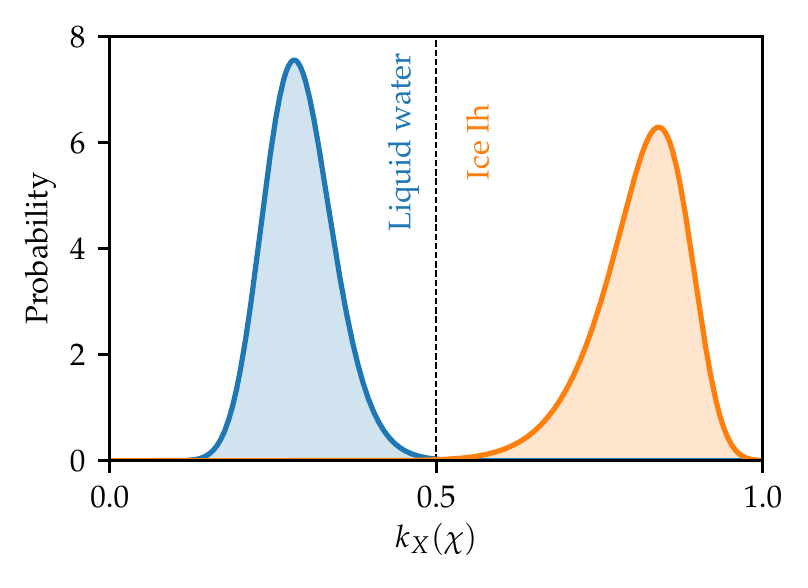}
\caption{\label{fig:Figure2} Distribution of the similarity kernel $k_X(\chi)$ in liquid water and in ice Ih. 
The distributions were calculated from simulations at 270 K and 1 bar using 288 water molecules.
}
\end{figure}

Since there is one value of $k_X(\chi)$ per oxygen atom, any given bulk configuration will have a distribution of this quantity.
We show in Fig.~\ref{fig:Figure2} the distributions of $k_X(\chi)$ in liquid water and in ice Ih at 270 K.
The two distributions have a small overlap and therefore $k_X(\chi)$ can be used to distinguish between liquid water and ice Ih environments.
Here we have chosen the spread of the Gaussians $\sigma$ such that the two distributions are approximately symmetrical with respect to $k_X(\chi) \approx 0.5$.
This will turn out useful below when we define a threshold between values of $k_X(\chi)$ consistent with the liquid and those consistent with the solid.
The rationale behind the choice of $\sigma$ is the following.
A large value of $\sigma$ gives a more lenient definition of the target environment and thus shifts the distributions to the right.
On the other hand, a small value of $\sigma$ gives a too strict definition and thermal motion of atoms will create too large a deviation from the target environments.
In this case the distributions are shifted to the left.
In between these extremes one finds a value of $\sigma$ that leads to distributions such as those shown in Fig.~\ref{fig:Figure2}.

The similarity kernel defined in Eq.\ \eqref{eq:kernel_multi1} provides a way to characterize the environments in a given configuration as being compatible with the environments in ice Ih.
For a system with $N$ water molecules there will be $N$ oxygen-oxygen environments $\chi^1,\chi^2,...,\chi^N$.
We shall define two global order parameters.
The first is the average value of the similarity kernel:
\begin{equation}
    \bar{k} = \frac{\sum_{i=1}^N k_X(\chi^i)}{N}, 
    \label{eq:op_avg}
\end{equation}
and the second is the number of environments consistent with the ice Ih environments, 
\begin{equation}
    n_{ice} = \{\mathrm{number \: of} \: \chi^i \: : \: k_X(\chi^i)>\kappa \}, 
    \label{eq:op_num}
\end{equation}
where $\kappa$ is a watershed between values of $k_X(\chi^i)$ consistent with the liquid and those consistent with the solid. 
According to our choice of $\sigma$ a reasonable value of $\kappa$ is $0.5$.
The order parameter defined in Eq.~\eqref{eq:op_num} has to be made continuous and differentiable to be used in enhanced sampling simulations.
We refer the reader to section \ref{sec:comp_details} for an appropriate definition of $n_{ice}$.
Equipped with these order parameters able to distinguish the liquid phase from ice Ih we shall now describe the enhanced sampling methodology. \\

\section{\label{sec:biaseddistrib}Biased distribution of the order parameter}

\begin{figure}
\includegraphics{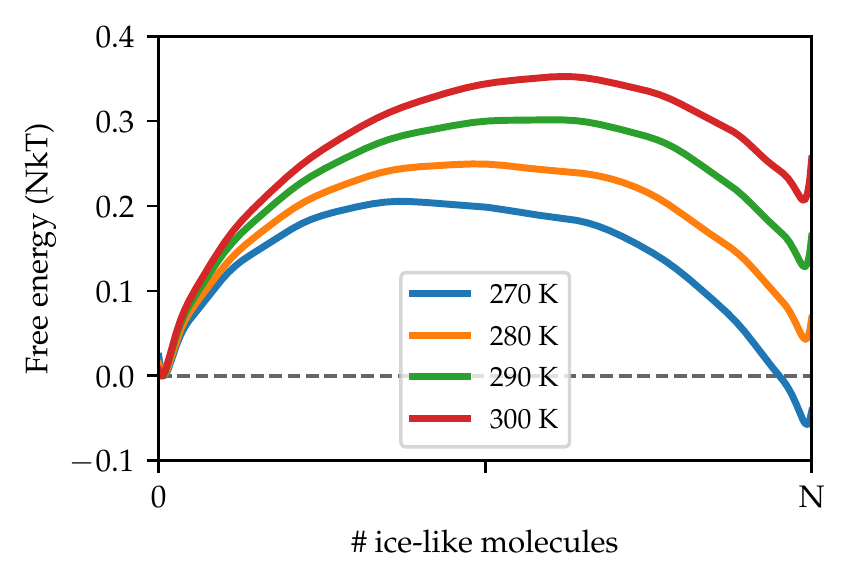}
\caption{\label{fig:Figure3} Free energy as a function of the number of ice-like molecules as defined by $n_{ice}$ for 4 different temperatures in the range 270-300 K.
A system of $N=96$ water molecules was used. 
The free energy is defined in Eq.\ \ref{eq:fes}.}
\end{figure}

As discussed in the introduction, first order phase transitions are characterized by a free energy barrier.
We anticipate some of the results and we plot in Fig.~\ref{fig:Figure3} the free energy as a function of the order parameter $n_{ice}$ defined as,
\begin{equation}
    G(n_{ice}) = -\frac{1}{\beta} \log p(n_{ice})
    \label{eq:fes}
\end{equation}
where the marginal probability $p(n_{ice})$ is,
\begin{equation}
    p(n_{ice}) = \int d\mathbf{R} d\mathcal{V} \: \frac{e^{-\beta [U(\mathbf{R},\mathcal{V}) + P\mathcal{V}]}}{Z_{\beta,P}} \: \delta(n_{ice}-n_{ice}(\mathbf{R})),
\end{equation}
$\beta$ is the inverse temperature, $P$ is the pressure, $\mathcal{V}$ is the volume, $U(\mathbf{R},\mathcal{V})$ is the potential energy, and $Z_{\beta,P}$ is the appropriate partition function.
For every temperature Fig.~\ref{fig:Figure3} shows two minima at $\sim 0$ and $\sim N$ that correspond to the liquid and ice Ih. 
Furthermore, it shows that the barrier for the transformation is around $\sim$ 25-30 kT and therefore a standard molecular dynamics simulation cannot provide ergodic sampling.
For this reason we aim at performing a simulation that samples a probability distribution different from that of the isothermal-isobaric ensemble.
We choose to sample the so called well-tempered distribution\cite{Bonomi10,Barducci08} of $n_{ice}$.
This distribution is defined as,
\begin{equation}
    p_{WT}(n_{ice}) \propto p(n_{ice})^{1/\gamma}
\end{equation}
where $\gamma>1$ is known as bias factor.
The effective free energy as a function of $n_{ice}$ is then,
\begin{equation}
    G_{WT}(n_{ice}) = G(n_{ice})/\gamma + C
    \label{eq:eff_fes}
\end{equation}
with $C$ an immaterial constant.
The free energy barrier is thus reduced by the factor $\gamma$ and we shall choose this parameter such that the barrier in $G_{WT}(n_{ice})$ is approximately $1$ kT.
Consequently, we achieve ergodic sampling of the liquid and solid phases.

In order to sample the well-tempered distribution we introduce a bias potential $V(n_{ice})$ that is a function of the order parameter.
We shall calculate $V(n_{ice})$ using a variational principle\cite{Valsson14} such that the sampled distribution is $p_{WT}(n_{ice})$.
During an initial stage of the simulation we will optimize a set of variational coefficients until the distribution $p_{WT}(n_{ice})$ is sampled. 
Once this target distribution is reached we will fix the variational parameters and continue the simulation with a static bias potential $V(n_{ice})$.
Further details can be found below in section \ref{sec:comp_details}.

The first simulations carried out with this approach resulted in the crystallization of structures with misorientation and stacking faults.
This had also been observed in ref.~\citenum{Piaggi19b} and we employ the same strategy used in that work to avoid these structures.
The strategy is based on the introduction of a bias potential that discourages these structures.
Details can be found in section \ref{sec:comp_details}.

\begin{figure*}
\includegraphics{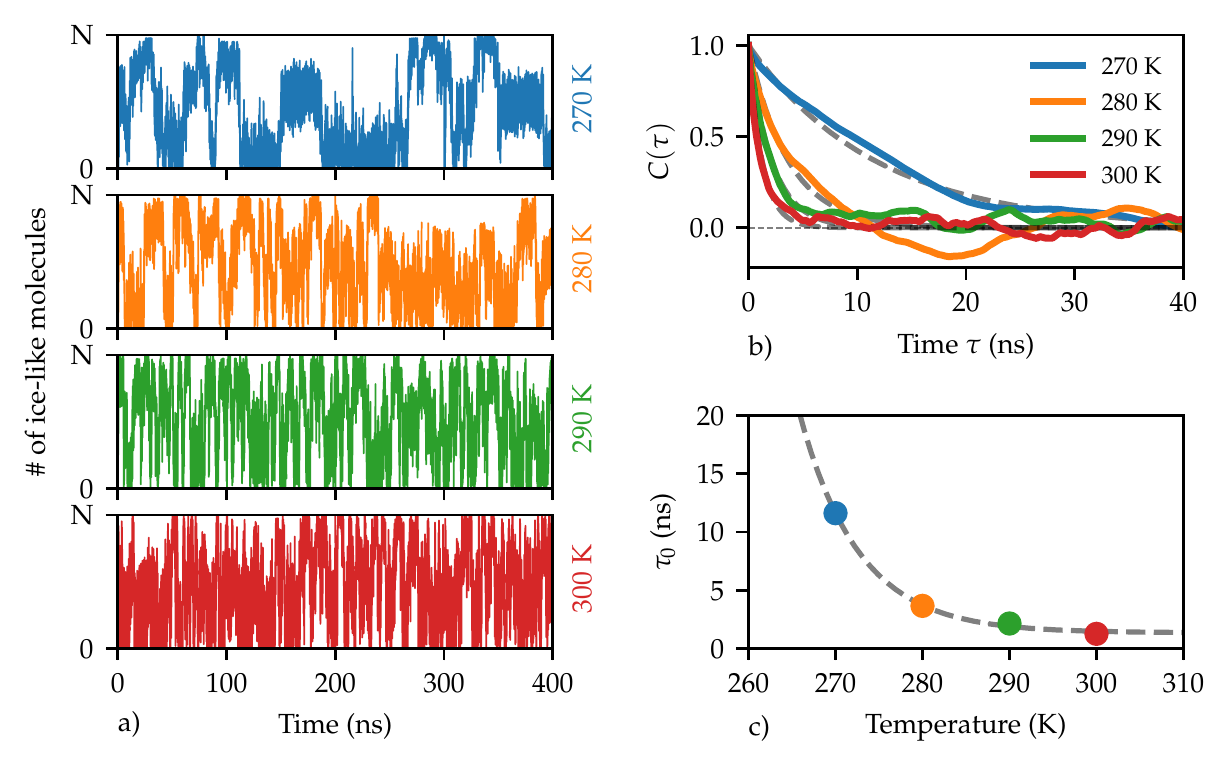}
\caption{\label{fig:Figure4} Dynamics of the liquid-ice Ih transformation under the action of a bias potential as a function of temperature. a) Number of ice-like molecules as defined by $n_{ice}$ (see text for details) as a function of simulation time for temperatures in the range 270-300 K. b) Time autocorrelation functions of $n_{ice}$ as defined in Eq.\ \eqref{eq:time_autocorr} and fits to the exponential decaying function $e^{-\tau/\tau_0}$. c) Characteristic correlation times $\tau_0$ as a function of temperature. The fit to an exponential function is included to guide the eyes. We stress that the correlation times reported here were calculated under the action of a static bias potential and that they are not correlation times of the crystallization/melting process.}
\end{figure*}

\section{\label{sec:free_energy} Calculation of free energy differences}

We are now in a position to calculate the free energy difference between ice Ih and the liquid $\Delta G_{l \rightarrow i}$ from the simulation described above.
$\Delta G_{l \rightarrow i}$ in the isothermal-isobaric ensemble at inverse temperature $\beta$ and pressure $P$ is defined as:
\begin{equation}
    \Delta G_{l \rightarrow i} = - \frac{1}{\beta} \log \left ( 
    \frac{\int\limits_{\mathrm{ice} \: \mathrm{Ih}} d\mathbf{R} \: d\mathcal{V} \: p(\mathbf{R},\mathcal{V}) }
    {\int\limits_{\mathrm{liquid}} d\mathbf{R} \: d\mathcal{V} \: p(\mathbf{R},\mathcal{V})}
    \right)
\end{equation}
where $p(\mathbf{R},\mathcal{V})=e^{-\beta [U(\mathbf{R},\mathcal{V}) + P\mathcal{V}]} / Z_{\beta,P}$.
$\Delta G_{l \rightarrow i}$ can be written using ensemble averages by introducing the Heaviside function,
\begin{equation}
    H(n_{ice}-n_{ice}*) = \begin{cases} 1 \: \text{if} \: n_{ice}>n_{ice}* \\  0 \: \text{if} \: n_{ice}<n_{ice}* \end{cases}
\end{equation}
where $n_{ice}*$ is some characteristic value of the order parameter that separates the liquid and ice Ih and thus,
\begin{equation}
  \Delta G_{l \rightarrow i} = - \frac{1}{\beta} \log \left( \frac{\langle H(n_{ice}-n_{ice}*) \rangle }{\langle 1-H(n_{ice}-n_{ice}*) \rangle }  \right ).
\end{equation}
Since the regions of high free energy do not contribute significantly to $\Delta G_{l \rightarrow i}$, the results of the calculations are not very sensitive to the choice of $n_{ice}*$.
A good and simple choice is $n_{ice}* = N/2$.

Since the simulation was not performed in the isothermal-isobaric ensemble but rather in a biased one, the equation above has to be rewritten in terms of ensemble averages in the latter ensemble.
Thus,
\begin{equation}
  \Delta G_{l \rightarrow i} = - \frac{1}{\beta} \log \left( \frac{\langle H(n_{ice}-n_{ice}*) w(\mathbf{R},\mathcal{V}) \rangle_V }{\langle 1-H(n_{ice}-n_{ice}*) w(\mathbf{R},\mathcal{V}) \rangle_V }  \right )
  \label{eq:free_energy_with_weights}
\end{equation}
where $w(\mathbf{R},\mathcal{V})=e^{\beta V(\mathbf{R},\mathcal{V})}$ are the weights associated to the bias potentials acting on the system, and $\langle \cdot \rangle_V$ denotes an ensemble average in the biased ensemble.
This is the equation that we shall use to calculate $\Delta G_{l \rightarrow i}$ from the simulations.

\section{\label{sec:results} Results}

We now turn to discuss the results of the calculations.
We performed simulations with three different system sizes composed of 16, 96, and 288 water molecules.
We studied temperatures from 270 K to 300 K based on previous reports of the melting temperature of ice Ih described by the TIP4P/Ice potential\cite{Garcia06,Abascal05}.
We show in Figure \ref{fig:Figure3} the free energy defined in Eq. \eqref{eq:fes} as a function of temperature for the system of 96 water molecules.
At 270 K ice Ih is more stable than the liquid but the stability is reversed at 280 K and above.
As previously mentioned, the free energy barriers are around $\sim$ 25-30 kT.

\begin{figure*}
\includegraphics{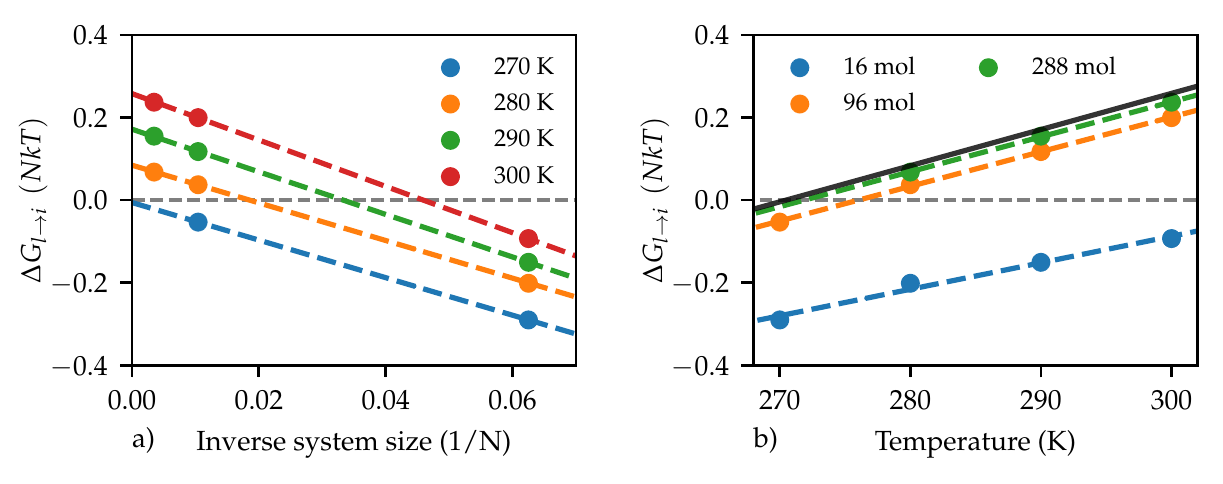}
\caption{\label{fig:Figure5} Free energy difference between liquid water and ice Ih $\Delta G_{l \rightarrow i}$. a) Inverse system size ($1/N$) vs. $\Delta G_{l \rightarrow i}$ is plotted to illustrate the finite size effects. A straight line was fit to the results at each temperature. The y-intercept of this line is considered to be the extrapolation to the large-$N$ limit. b) Temperature vs. $\Delta G_{l \rightarrow i}$. The black line corresponds to the large-$N$ limit and the x-intercept is the melting temperature $T_m \sim 270 K$. The dashed lines are fits to the data for different $N$. }
\end{figure*}

In Fig.~\ref{fig:Figure4}a we plot $n_{ice}$ as a function of simulation time at different temperatures.
The figure shows that the number of transitions from the liquid to the solid per unit time decreases as the temperature is lowered.
In order to quantify this effect we calculated for each temperature the time autocorrelation function of $n_{ice}$,
\begin{equation}
    C(\tau) = \frac{\langle \widetilde{n_{ice}}(t+\tau) \widetilde{n_{ice}}(t) \rangle_V}{\langle \widetilde{n_{ice}}(t)^2\rangle_V}
    \label{eq:time_autocorr}
\end{equation}
where $\widetilde{n_{ice}}(t)=n_{ice}(t)-\langle n_{ice}(t)\rangle_V$ and we have used $\langle \cdot \rangle_V$ to emphasize that the average was done using the biased trajectories.
We plot $C(\tau)$ in Figure ~\ref{fig:Figure4}b.
We also fitted a decaying exponential function $e^{-\tau/\tau_0}$ to $C(\tau)$ in order to calculate a characteristic correlation time $\tau_0$.
Figure ~\ref{fig:Figure4}c shows $\tau_0$ as a function of temperature and from this plot we see that the autocorrelation time increases exponentially as the temperature is lowered.
Note that these are not the system's physical autocorrelation times but those of the system under the influence of the bias potential.

The phenomenon described in the previous paragraph bears some resemblance to the critical slowing down found in Monte Carlo simulations of lattice models and typically characterized with a dynamical exponent\cite{NewmanBarkemaBook}.
However, the situation here is different since the slowing down is most likely a result of a rough free energy landscape in directions orthogonal to our order parameter that gives rise to glass-like behavior.
We ruled out the possibility that this behavior is a consequence of lack of convergence of the bias potential by plotting the biased free energy as a function of $n_{ice}$ for each temperature.
The barriers found ranged from 1.5 to 3 kT and therefore cannot be responsible of the behavior shown in Fig.~\ref{fig:Figure4}.
Furthermore, this phenomenon has also been observed for another glass former, namely silica\cite{Niu18}.
A possible way to address this issue is the parallel tempering\cite{Hansmann97,Sugita99} technique or the approach outlined in ref.\ \citenum{Yang18}.

We now set out to study the finite size effects in the melting temperature of ice Ih.
For this purpose we repeated the calculations described above using 16 and 288 water molecules instead of 96.
Then, we calculated the free energy differences $\Delta G_{l \rightarrow i}$ using Eq.\ \eqref{eq:free_energy_with_weights}.
The results thus obtained are shown in Fig.~\ref{fig:Figure5}a where we plot $\Delta G_{l \rightarrow i}$ vs. inverse system size $1/N$ at four different temperatures.
We also report $\Delta G_{l \rightarrow i}$ in Table \ref{tab:table1}.
From the data at each temperature we extrapolated the results to $N\to\infty$.
$\Delta G_{l \rightarrow i}$ scales linearly with $1/N$ as predicted by the theory of finite size scaling in first order phase transitions\cite{Binder87}.

We then proceed to plot $\Delta G_{l \rightarrow i}$ vs. temperature in Fig.~\ref{fig:Figure5}b including the results of the extrapolation to the thermodynamic limit described in the previous paragraph.
By fitting a straight line to the results in the thermodynamic limit we obtain our best estimate of the melting temperature which is $T_m \sim 270 K$.
We summarize in Table \ref{tab:table2} the results obtained using different methods.

\begin{table}[b]
\caption{\label{tab:table1} Differences in free energy between ice Ih and the liquid $\Delta G_{l \rightarrow i}$ as a function of temperature and system size. The errors shown in parentheses were calculated using block averages.}
\begin{ruledtabular}
\begin{tabular}{c c | c c}
 Temperature (K) & $\Delta G_{l \rightarrow i}$ (NkT) &  Temperature (K) & $\Delta G_{l \rightarrow i}$ (NkT) \\
 \hline
 16 molecules &  & 288 molecules & \\
 270 & -0.29(3) &   &  \\
 280 & -0.20(3) & 280 & 0.068(1) \\
 290 & -0.150(8) & 290 & 0.15(1) \\
 300 & -0.093(5) & 300 & 0.24(1) \\
 \hline
  96 molecules & & $\infty$ molecules & \\
 270 & -0.053(5) & 270 & -0.006(8) \\
 280 & 0.037(5)  & 280 & 0.084(7) \\
 290 & 0.123(5)  & 290 & 0.175(5) \\
 300 & 0.200(5)  & 300 & 0.257(5) \\
\end{tabular}
\end{ruledtabular}
\end{table}

\begin{table}[b]
\caption{\label{tab:table2} Melting temperature of ice Ih described by the TIP4P/Ice potential obtained with different methods. The error of the melting temperature is calculated from the standard deviation of the parameters in the linear fit of $\Delta G_{l \rightarrow i}$.}
\begin{ruledtabular}
\begin{tabular}{cc}
 Method & Melting temperature (K) \\
 \hline
 This work & 270(2) \\
 Hamiltonian Gibbs-Duhem integration \cite{Vega05,Abascal05} & 272(6) \\
 Coexistence\cite{Garcia06} & 268(2) \\
 Coexistence\cite{Conde17} & 269.8(1) \\
 Free surface\cite{Vega06} & 271(1) \\
 Experimental & 273.15 \\
\end{tabular}
\end{ruledtabular}
\end{table}

\begin{figure}
\includegraphics{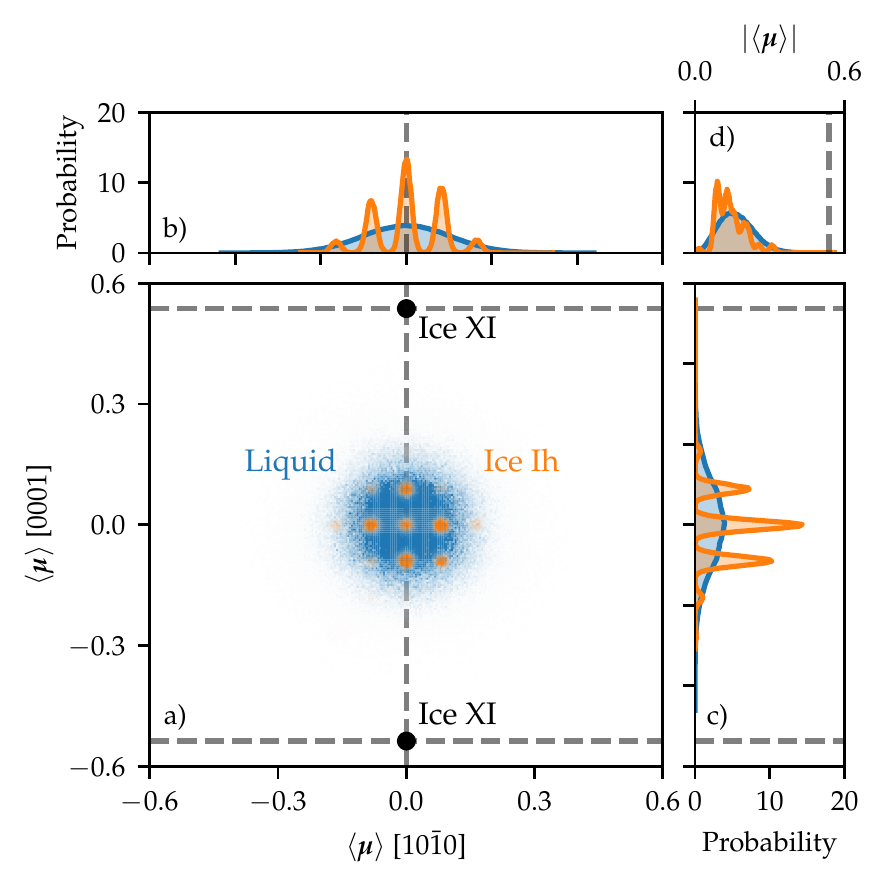}
\caption{\label{fig:Figure6} Analysis of the proton disorder using the average dipole moment direction $\boldsymbol{\mu} = \sum_{i=1}^N \hat{\boldsymbol\mu}_i /N $ where $\hat{\boldsymbol\mu}_i$ is the dipole moment versor of molecule $i$ and $N$ is the number of molecules. a) 2D histogram of the projection of $\boldsymbol{\mu}$ along two directions. 
The x-axis is the direction perpendicular to the prismatic plane $(10\bar{1}0)$ and the y-axis is the direction perpendicular to the basal plane $(0001)$.
b) and c) 1D histograms along the same directions described above.
d) Histogram of the norm of $\boldsymbol{\mu}$ denoted with $|\boldsymbol{\mu}|$.
Results for liquid water and ice Ih are shown in blue and orange, respectively.
The data was gathered from the simulation using $N=96$ water molecules at 300 K.
$\boldsymbol{\mu}$ for the proton ordered phase, i.e. ice XI, is shown as black dots and with black dashed lines.
}
\end{figure}

The agreement between our method and thermodynamic integration\cite{Vega05,Abascal05} is remarkable if one takes into account that the latter requires as input the entropic contribution of the proton disorder.
Instead, in our method the proton configuration appears spontaneously during crystallization and there is no explicit bias because the order parameter is only a function of the positions of oxygen atoms.
Below we analyze the proton configurations obtained from our simulations.

The proton configuration will determine the total dipole moment of the system.
We will thus use the dipole moment per molecule, $\boldsymbol{\mu} = \sum_{i=1}^N \hat{\boldsymbol\mu}_i /N $, where $\hat{\boldsymbol\mu}_i$ is the dipole moment versor of molecule $i$ and $N$ is the number of molecules, to analyze the proton configuration.
We note that in the equation above we assume the dipole moment of each molecule to be one.
Since the molecules are rigid, using a different value would not change our results qualitatively.
In Fig.~\ref{fig:Figure6}a we plot a 2D histogram of the projection of $\boldsymbol{\mu}$ along two directions. 
The x-axis is the direction perpendicular to the prismatic plane $(10\bar{1}0)$ and the y-axis is the direction perpendicular to the basal plane $(0001)$.
We also show in Figs.~\ref{fig:Figure6}b and c the 1D histograms along the same directions described in Fig.~\ref{fig:Figure6}a.
In these Figures we compare the results for liquid water (blue) and ice Ih (orange) for a system of $N=96$ water molecules at 300 K.
Both distributions are centered at $\boldsymbol{\mu}=\boldsymbol{0}$ as expected for non-ferroelectric phases like liquid water and ice Ih.
Therefore our configurations are compatible with the criterion used by Rahman and Stillinger to construct proton disordered configurations\cite{Rahman72}.
The distribution of $\boldsymbol{\mu}$ in water can be well described by a multivariate normal distribution.
On the other hand, the distribution of $\boldsymbol{\mu}$ in ice Ih shows several peaks that form a lattice.
This is expected since the crystal structure of the oxygen atoms allows only a restricted number of directions for the dipole moment (6 in each environment).

In Fig.~\ref{fig:Figure6} we also show $\boldsymbol{\mu}$ for the proton ordered phase, i.e. ice XI.
The configurations obtained during the simulations are far away from the ordered configurations and are also centered at $\boldsymbol{\mu}=\boldsymbol{0}$.
This suggests that our configurations are representative of the proton disorder in ice Ih.

We also plot the distribution of the norm $|\boldsymbol{\mu}|$ in Fig.~\ref{fig:Figure6}d.
The distribution of the liquid is very similar to the Maxwell-Boltzmann distribution as follows from the fact that $\boldsymbol{\mu}$ has a multivariate normal distribution.
The distribution of ice Ih also resembles the  Maxwell-Boltzmann distribution although there are some anomalous peaks.
Rahman and Stillinger\cite{Rahman72} found a smoother distribution for a larger system and it is reasonable to think that in the thermodynamic limit the multiple peaks in our histogram would also be absent.
Indeed, we repeated our analysis for the system with $n=288$ molecules and found an increase in the number of peaks that are also closer to each other.

\section{\label{sec:conclusions} Conclusions}

We have calculated the melting temperature of ice Ih as described by the TIP4P/Ice water model using enhanced sampling molecular simulations.
Our best estimate of the melting temperature is 270 K in the thermodynamic limit.
This result agrees with previous estimates that put the melting temperature between 268 and 272 K.
Therefore the results show that enhanced sampling simulations offer a state-of-the-art alternative to calculating melting temperatures and differences in free energy between the liquid and the solid.

A key feature of this approach is the use of an order parameter that enforces a particular orientation of the crystal structure.
The order parameter is based on a measure of the similarity between the atomic environments in a given configuration and reference environments.
The reference environments were chosen in order to obtain the crystal structure of ice Ih and included 17 nearest neighbors.

We argue that there are several advantages to our approach.
For one, since the crystallization and melting processes are explicitly simulated, physical insight can be extracted directly from these simulations.
For instance, we could initially observe stacking faults that point to the competition between ice Ih and ice Ic. 
We later improved our simulation setup in order to avoid this phenomenon and obtained only ice Ih.
In addition, our results automatically include the effects of proton disorder.
Our order parameter does not include any information about the orientation of the water molecules and therefore it does not bias the structure towards any particular proton configuration.

We also observed an exponentially slower sampling as the temperature is decreased.
This is surprising if one considers that the free energy as a function of the order parameter has been flattened.
We suggest that this result is a consequence of a rough free energy landscape in directions orthogonal to our order parameter.
This observation would be compatible with the tendency of water to form glasses.

\section{\label{sec:comp_details} Computational details}

The simulations were carried out using \textsc{LAMMPS}\cite{Plimpton95} patched with the \textsc{PLUMED} 2\cite{Tribello14} enhanced sampling plugin.
\textsc{PLUMED} 2 was supplemented with the VES module\cite{vescode}.
The input files to reproduce all simulations are available on the \textsc{PLUMED-NEST}\cite{Bonomi19} as part of a collective effort to improve the transparency and reproducibility of enhanced molecular simulations.
We employed a timestep for the integration of the equations of motion of 2 fs in all simulations.
The temperature was controlled using the stochastic velocity rescaling algorithm\cite{Bussi07} with a 0.1 ps relaxation time.
The pressure was maintained at 1 bar using an isotropic version of the Parrinello-Rahman barostat\cite{Parrinello81} with a 1 ps relaxation time.
The bond lengths and angles were kept fixed using the \textsc{SHAKE} algorithm\cite{Ryckaert77}.
A cutoff of 0.85 nm was used for the Lennard-Jones and Coulomb interactions.
Long-range Coulomb interactions beyond this cutoff were computed with the particle-particle particle-mesh (PPPM) solver\cite{Hockney88}.
Tail corrections to the pressure and energy were included to take into account long-range effects neglected due to the Lennard-Jones potential truncation\cite{FrenkelBook}.

The bias potential was constructed using the variational principle of Valsson and Parrinello\cite{Valsson14} that we summarize below.
Within this formalism, the bias potential is determined through the minimization of the functional,
\begin{align}
\label{omega1}
\Omega [V] & =
\frac{1}{\beta} \log
\frac
{\int d\mathbf{s} \, e^{-\beta \left[ F(\mathbf{s}) + V(\mathbf{s})\right]}}
{\int d\mathbf{s} \, e^{-\beta F(\mathbf{s})}}
+
\int d\mathbf{s} \, p(\mathbf{s}) V(\mathbf{s}),
\end{align}
where $\mathbf{s}$ is a set of collective variables (CVs) that are a function of the atomic coordinates $\mathbf{R}$, the free energy is given within an immaterial constant by $F(\mathbf{s})=-\frac{1}{\beta}\log\int d\mathbf{R} \delta(\mathbf{s}-\mathbf{s}(\mathbf{R})) e^{-\beta U(\mathbf{R})}$,  $U(\mathbf{R})$ is the interatomic potential, and $p(\mathbf{s})$ is a preassigned target distribution.
The minimum of this convex functional is reached for:
\begin{equation}
\label{eq:optimal_bias}
V(\mathbf{s}) = -F(\mathbf{s})-{\frac {1}{\beta}} \log {p(\mathbf{s})}.
\end{equation}
which amounts to saying that in a system biased by $V(\mathbf{s})$, the distribution of the CVs is $p(\mathbf{s})$.

As described in the main part, we used $n_{ice}$ as CV and we targeted the so called well-tempered distribution\cite{Valsson15}, i.e. $p(n_{ice}) \propto P(n_{ice})^{1/\gamma}$.
We employed bias factors $\gamma$ of 30, 50, and 100 for the systems of 16, 96, and 288 molecules, respectively.
The bias potential $V(\mathbf{s})$ was expanded in Legendre polynomials of order 20 defined in the interval $[0,N]$.
The functional $\Omega [V]$ was minimized using the averaged stochastic gradient descent algorithm\cite{Valsson14,Bach13} and the gradient was averaged over 500 timesteps.
The step size in the optimization was 2 kJ/mol.
The well-tempered distribution was determined self-consistently as described in ref.~\citenum{Valsson15} with update frequency 100 optimization iterations.
Four multiple walkers\cite{Raiteri06} contributed to the statistics to calculate the gradient of $\Omega [V]$.
Once the bias potential was deemed converged the optimization was stopped and the simulation was continued with a static bias potential akin to the ones employed in the umbrella sampling technique\cite{Torrie77}.
All the reported quantities were calculated under the action of a static bias potential.

The order parameters in Eqs.~\eqref{eq:op_avg} and \eqref{eq:op_num} require the specification of the environments $X$ and the spread of the Gaussians $\sigma$.
We employed $\sigma=0.076$, $0.055$, $0.055$ for the systems of $N=16$ ,$96$, $288$ molecules, respectively.
In Eq.~\eqref{eq:op_num} we defined an order parameter $n_{ice}$ that counts the number of molecules that satisfy $k_{X}(\chi^i)>\kappa$ where $\kappa$ is chosen to be $0.5$.
However, the definition given above cannot be used in enhanced sampling simulations since it is not continuous and differentiable.
For this reason in the simulations we employed the following expression\cite{Tribello17},
\begin{equation}
n_{ice}=N-\sum\limits_{i=1}^{N}  \frac{1-(k_{X}(\chi^i)/\kappa)^{p}}{1-(k_{X}(\chi^i)/\kappa)^{q}}
\label{eq:op_num_cd}
\end{equation}
with $p=15$ and $q=30$ for the 96 molecule case, and $p=12$ and $q=24$ for the 96 and 288 molecule cases.
In the limit of large $q$ and $p$ Eqs.~\eqref{eq:op_num} and \eqref{eq:op_num_cd} yield the same result.
By the same token, Eq.~\eqref{eq:kernel_multi1} cannot be used in enhanced sampling simulations.
A continuous and differentiable variant of Eq.~\eqref{eq:kernel_multi1} is,
\begin{equation}
 k_X(\chi)= \frac{1}{\lambda} \log \left ( \sum\limits_{l=1}^{4}\exp \left (\lambda \: k_{\chi_l}(\chi) \right ) \right ),
 \label{eq:kernel_multi1_cd}
\end{equation}
where $\lambda=100$ was chosen, and the index $l$ runs over the 4 local environments of ice Ih.
This is the expression we have used in the simulations.
For $\lambda \to \infty$, Eq.~\eqref{eq:kernel_multi1_cd} selects the largest $k_{\chi_l}(\chi)$ with $\chi_l \in X$.

As described above, we added bias potentials to discourage the formation of structures with misorientation or stacking faults.
For this purpose we defined two collective variables.
The first aims at distinguishing structures misaligned with respect to the box as is defined as,
\begin{equation}
s_{c}^1 = \frac{Q_6-Q_6^l}{Q_6^i-Q_6^l} - \frac{\bar{k}-\bar{k}^l}{\bar{k}^i-\bar{k}^l}
\label{eq:constraint1}
\end{equation}
where $Q_6$ is the global Steinhardt parameter\cite{Steinhardt83} as defined in ref.\ \citenum{vanDuijneveldt92} and $\bar{k}$ is defined in Eq.~\eqref{eq:op_avg}.
The superscripts in Eq.~(\ref{eq:constraint1}) refer to the values of the order parameters in the liquid ($l$) and in ice Ih ($i$).
The rationale behind $s_{c}^1$ has been described in ref.~\citenum{Piaggi19b}.
On the other hand, the purpose of the second collective variable is to distinguish perfect structures from those with stacking faults.
It is defined as,
\begin{equation}
s_{c}^2 = \frac{\bar{k_c}-\bar{k_c}^l}{\bar{k_c}^i-\bar{k_c}^l} - \frac{\bar{k}-\bar{k}^l}{\bar{k}^i-\bar{k}^l}
\label{eq:constraint}
\end{equation}
where the symbols with the subscript $c$ refer to the kernel defined for the cubic diamond structure.

In the simulations $s_{c}^1$ and/or $s_{c}^2$ were restrained with harmonic potentials,
\begin{equation}
V(s_c^{\alpha})= 
\begin{cases}
\kappa (s_c^{\alpha}-\widetilde{s_c^{\alpha}})^2 \quad & \mathrm{if} \quad  s_c^{\alpha}>\widetilde{s_c^{\alpha}} \\
0 \quad & \mathrm{otherwise}
\end{cases}
\end{equation}
and the chosen $c$ and $\widetilde{s_c^{\alpha}}$ for each simulation can be found in the simulation input files made available.

\begin{acknowledgments}
P.M.P thanks Marcos Calegari for useful discussions and Zachary Goldsmith for carefully reading the manuscript.
P.M.P was supported by an Early Postdoc.Mobility fellowship from the Swiss National Science Foundation. 
This work was conducted within the center: Chemistry in Solution and at Interfaces funded by the DoE under Award DE-SC0019394.
The calculations reported in this work were performed using the Princeton Research Computing resources at Princeton University.
\end{acknowledgments}

\section*{Data Availability Statement}
The input files and results of the simulations are openly available on GitHub\cite{GitHubRepo}, and on \textsc{PLUMED-NEST} (www.plumed-nest.org), the public repository of the PLUMED consortium\cite{Bonomi19}, as plumID:20.010.

%
%


\begin{thebibliography}{42}%
\makeatletter
\providecommand \@ifxundefined [1]{%
 \@ifx{#1\undefined}
}%
\providecommand \@ifnum [1]{%
 \ifnum #1\expandafter \@firstoftwo
 \else \expandafter \@secondoftwo
 \fi
}%
\providecommand \@ifx [1]{%
 \ifx #1\expandafter \@firstoftwo
 \else \expandafter \@secondoftwo
 \fi
}%
\providecommand \natexlab [1]{#1}%
\providecommand \enquote  [1]{``#1''}%
\providecommand \bibnamefont  [1]{#1}%
\providecommand \bibfnamefont [1]{#1}%
\providecommand \citenamefont [1]{#1}%
\providecommand \href@noop [0]{\@secondoftwo}%
\providecommand \href [0]{\begingroup \@sanitize@url \@href}%
\providecommand \@href[1]{\@@startlink{#1}\@@href}%
\providecommand \@@href[1]{\endgroup#1\@@endlink}%
\providecommand \@sanitize@url [0]{\catcode `\\12\catcode `\$12\catcode
  `\&12\catcode `\#12\catcode `\^12\catcode `\_12\catcode `\%12\relax}%
\providecommand \@@startlink[1]{}%
\providecommand \@@endlink[0]{}%
\providecommand \url  [0]{\begingroup\@sanitize@url \@url }%
\providecommand \@url [1]{\endgroup\@href {#1}{\urlprefix }}%
\providecommand \urlprefix  [0]{URL }%
\providecommand \Eprint [0]{\href }%
\providecommand \doibase [0]{http://dx.doi.org/}%
\providecommand \selectlanguage [0]{\@gobble}%
\providecommand \bibinfo  [0]{\@secondoftwo}%
\providecommand \bibfield  [0]{\@secondoftwo}%
\providecommand \translation [1]{[#1]}%
\providecommand \BibitemOpen [0]{}%
\providecommand \bibitemStop [0]{}%
\providecommand \bibitemNoStop [0]{.\EOS\space}%
\providecommand \EOS [0]{\spacefactor3000\relax}%
\providecommand \BibitemShut  [1]{\csname bibitem#1\endcsname}%
\let\auto@bib@innerbib\@empty
\bibitem [{\citenamefont {Sanz}\ \emph {et~al.}(2004)\citenamefont {Sanz},
  \citenamefont {Vega}, \citenamefont {Abascal},\ and\ \citenamefont
  {MacDowell}}]{Sanz04}%
  \BibitemOpen
  \bibfield  {author} {\bibinfo {author} {\bibfnamefont {E.}~\bibnamefont
  {Sanz}}, \bibinfo {author} {\bibfnamefont {C.}~\bibnamefont {Vega}}, \bibinfo
  {author} {\bibfnamefont {J.}~\bibnamefont {Abascal}}, \ and\ \bibinfo
  {author} {\bibfnamefont {L.}~\bibnamefont {MacDowell}},\ }\bibfield  {title}
  {\enquote {\bibinfo {title} {Phase diagram of water from computer
  simulation},}\ }\href@noop {} {\bibfield  {journal} {\bibinfo  {journal}
  {Physical review letters}\ }\textbf {\bibinfo {volume} {92}},\ \bibinfo
  {pages} {255701} (\bibinfo {year} {2004})}\BibitemShut {NoStop}%
\bibitem [{\citenamefont {Vega}, \citenamefont {Sanz},\ and\ \citenamefont
  {Abascal}(2005)}]{Vega05}%
  \BibitemOpen
  \bibfield  {author} {\bibinfo {author} {\bibfnamefont {C.}~\bibnamefont
  {Vega}}, \bibinfo {author} {\bibfnamefont {E.}~\bibnamefont {Sanz}}, \ and\
  \bibinfo {author} {\bibfnamefont {J.}~\bibnamefont {Abascal}},\ }\bibfield
  {title} {\enquote {\bibinfo {title} {The melting temperature of the most
  common models of water},}\ }\href@noop {} {\bibfield  {journal} {\bibinfo
  {journal} {The Journal of chemical physics}\ }\textbf {\bibinfo {volume}
  {122}},\ \bibinfo {pages} {114507} (\bibinfo {year} {2005})}\BibitemShut
  {NoStop}%
\bibitem [{\citenamefont {Garc{\'\i}a~Fern{\'a}ndez}, \citenamefont {Abascal},\
  and\ \citenamefont {Vega}(2006)}]{Garcia06}%
  \BibitemOpen
  \bibfield  {author} {\bibinfo {author} {\bibfnamefont {R.}~\bibnamefont
  {Garc{\'\i}a~Fern{\'a}ndez}}, \bibinfo {author} {\bibfnamefont {J.~L.}\
  \bibnamefont {Abascal}}, \ and\ \bibinfo {author} {\bibfnamefont
  {C.}~\bibnamefont {Vega}},\ }\bibfield  {title} {\enquote {\bibinfo {title}
  {The melting point of ice {I}h for common water models calculated from direct
  coexistence of the solid-liquid interface},}\ }\href@noop {} {\bibfield
  {journal} {\bibinfo  {journal} {The Journal of chemical physics}\ }\textbf
  {\bibinfo {volume} {124}},\ \bibinfo {pages} {144506} (\bibinfo {year}
  {2006})}\BibitemShut {NoStop}%
\bibitem [{\citenamefont {Conde}, \citenamefont {Rovere},\ and\ \citenamefont
  {Gallo}(2017)}]{Conde17}%
  \BibitemOpen
  \bibfield  {author} {\bibinfo {author} {\bibfnamefont {M.}~\bibnamefont
  {Conde}}, \bibinfo {author} {\bibfnamefont {M.}~\bibnamefont {Rovere}}, \
  and\ \bibinfo {author} {\bibfnamefont {P.}~\bibnamefont {Gallo}},\ }\bibfield
   {title} {\enquote {\bibinfo {title} {High precision determination of the
  melting points of water tip4p/2005 and water tip4p/ice models by the direct
  coexistence technique},}\ }\href@noop {} {\bibfield  {journal} {\bibinfo
  {journal} {The Journal of chemical physics}\ }\textbf {\bibinfo {volume}
  {147}},\ \bibinfo {pages} {244506} (\bibinfo {year} {2017})}\BibitemShut
  {NoStop}%
\bibitem [{\citenamefont {Pedersen}\ \emph {et~al.}(2013)\citenamefont
  {Pedersen}, \citenamefont {Hummel}, \citenamefont {Kresse}, \citenamefont
  {Kahl},\ and\ \citenamefont {Dellago}}]{Pedersen13}%
  \BibitemOpen
  \bibfield  {author} {\bibinfo {author} {\bibfnamefont {U.~R.}\ \bibnamefont
  {Pedersen}}, \bibinfo {author} {\bibfnamefont {F.}~\bibnamefont {Hummel}},
  \bibinfo {author} {\bibfnamefont {G.}~\bibnamefont {Kresse}}, \bibinfo
  {author} {\bibfnamefont {G.}~\bibnamefont {Kahl}}, \ and\ \bibinfo {author}
  {\bibfnamefont {C.}~\bibnamefont {Dellago}},\ }\bibfield  {title} {\enquote
  {\bibinfo {title} {Computing gibbs free energy differences by interface
  pinning},}\ }\href@noop {} {\bibfield  {journal} {\bibinfo  {journal}
  {Physical Review B}\ }\textbf {\bibinfo {volume} {88}},\ \bibinfo {pages}
  {094101} (\bibinfo {year} {2013})}\BibitemShut {NoStop}%
\bibitem [{\citenamefont {Cheng}\ \emph {et~al.}(2019)\citenamefont {Cheng},
  \citenamefont {Engel}, \citenamefont {Behler}, \citenamefont {Dellago},\ and\
  \citenamefont {Ceriotti}}]{Cheng19}%
  \BibitemOpen
  \bibfield  {author} {\bibinfo {author} {\bibfnamefont {B.}~\bibnamefont
  {Cheng}}, \bibinfo {author} {\bibfnamefont {E.~A.}\ \bibnamefont {Engel}},
  \bibinfo {author} {\bibfnamefont {J.}~\bibnamefont {Behler}}, \bibinfo
  {author} {\bibfnamefont {C.}~\bibnamefont {Dellago}}, \ and\ \bibinfo
  {author} {\bibfnamefont {M.}~\bibnamefont {Ceriotti}},\ }\bibfield  {title}
  {\enquote {\bibinfo {title} {Ab initio thermodynamics of liquid and solid
  water},}\ }\href@noop {} {\bibfield  {journal} {\bibinfo  {journal}
  {Proceedings of the National Academy of Sciences}\ }\textbf {\bibinfo
  {volume} {116}},\ \bibinfo {pages} {1110--1115} (\bibinfo {year}
  {2019})}\BibitemShut {NoStop}%
\bibitem [{\citenamefont {Abascal}\ \emph {et~al.}(2005)\citenamefont
  {Abascal}, \citenamefont {Sanz}, \citenamefont {Garc{\'\i}a~Fern{\'a}ndez},\
  and\ \citenamefont {Vega}}]{Abascal05}%
  \BibitemOpen
  \bibfield  {author} {\bibinfo {author} {\bibfnamefont {J.}~\bibnamefont
  {Abascal}}, \bibinfo {author} {\bibfnamefont {E.}~\bibnamefont {Sanz}},
  \bibinfo {author} {\bibfnamefont {R.}~\bibnamefont
  {Garc{\'\i}a~Fern{\'a}ndez}}, \ and\ \bibinfo {author} {\bibfnamefont
  {C.}~\bibnamefont {Vega}},\ }\bibfield  {title} {\enquote {\bibinfo {title}
  {A potential model for the study of ices and amorphous water: Tip4p/ice},}\
  }\href {\doibase 10.1063/1.1931662} {\bibfield  {journal} {\bibinfo
  {journal} {The Journal of chemical physics}\ }\textbf {\bibinfo {volume}
  {122}},\ \bibinfo {pages} {234511} (\bibinfo {year} {2005})}\BibitemShut
  {NoStop}%
\bibitem [{\citenamefont {Valsson}\ and\ \citenamefont
  {Parrinello}(2014)}]{Valsson14}%
  \BibitemOpen
  \bibfield  {author} {\bibinfo {author} {\bibfnamefont {O.}~\bibnamefont
  {Valsson}}\ and\ \bibinfo {author} {\bibfnamefont {M.}~\bibnamefont
  {Parrinello}},\ }\bibfield  {title} {\enquote {\bibinfo {title} {Variational
  approach to enhanced sampling and free energy calculations},}\ }\href
  {\doibase 10.1103/PhysRevLett.113.090601} {\bibfield  {journal} {\bibinfo
  {journal} {Physical review letters}\ }\textbf {\bibinfo {volume} {113}},\
  \bibinfo {pages} {090601} (\bibinfo {year} {2014})}\BibitemShut {NoStop}%
\bibitem [{\citenamefont {Piaggi}\ and\ \citenamefont
  {Parrinello}(2019)}]{Piaggi19b}%
  \BibitemOpen
  \bibfield  {author} {\bibinfo {author} {\bibfnamefont {P.~M.}\ \bibnamefont
  {Piaggi}}\ and\ \bibinfo {author} {\bibfnamefont {M.}~\bibnamefont
  {Parrinello}},\ }\bibfield  {title} {\enquote {\bibinfo {title} {Calculation
  of phase diagrams in the multithermal-multibaric ensemble},}\ }\href@noop {}
  {\bibfield  {journal} {\bibinfo  {journal} {The Journal of chemical physics}\
  }\textbf {\bibinfo {volume} {150}},\ \bibinfo {pages} {244119} (\bibinfo
  {year} {2019})}\BibitemShut {NoStop}%
\bibitem [{\citenamefont {Stukowski}(2009)}]{Stukowski09}%
  \BibitemOpen
  \bibfield  {author} {\bibinfo {author} {\bibfnamefont {A.}~\bibnamefont
  {Stukowski}},\ }\bibfield  {title} {\enquote {\bibinfo {title} {Visualization
  and analysis of atomistic simulation data with ovito--the open visualization
  tool},}\ }\href {\doibase 10.1088/0965-0393/18/1/015012} {\bibfield
  {journal} {\bibinfo  {journal} {Modelling and Simulation in Materials Science
  and Engineering}\ }\textbf {\bibinfo {volume} {18}},\ \bibinfo {pages}
  {015012} (\bibinfo {year} {2009})}\BibitemShut {NoStop}%
\bibitem [{\citenamefont {Petrenko}\ and\ \citenamefont
  {Whitworth}(1999)}]{PetrenkoIce}%
  \BibitemOpen
  \bibfield  {author} {\bibinfo {author} {\bibfnamefont {V.~F.}\ \bibnamefont
  {Petrenko}}\ and\ \bibinfo {author} {\bibfnamefont {R.~W.}\ \bibnamefont
  {Whitworth}},\ }\href@noop {} {\emph {\bibinfo {title} {Physics of ice}}}\
  (\bibinfo  {publisher} {OUP Oxford},\ \bibinfo {year} {1999})\BibitemShut
  {NoStop}%
\bibitem [{\citenamefont {Bernal}\ and\ \citenamefont
  {Fowler}(1933)}]{Bernal33}%
  \BibitemOpen
  \bibfield  {author} {\bibinfo {author} {\bibfnamefont {J.~D.}\ \bibnamefont
  {Bernal}}\ and\ \bibinfo {author} {\bibfnamefont {R.~H.}\ \bibnamefont
  {Fowler}},\ }\bibfield  {title} {\enquote {\bibinfo {title} {A theory of
  water and ionic solution, with particular reference to hydrogen and hydroxyl
  ions},}\ }\href@noop {} {\bibfield  {journal} {\bibinfo  {journal} {The
  Journal of Chemical Physics}\ }\textbf {\bibinfo {volume} {1}},\ \bibinfo
  {pages} {515--548} (\bibinfo {year} {1933})}\BibitemShut {NoStop}%
\bibitem [{\citenamefont {Pauling}(1935)}]{Pauling35}%
  \BibitemOpen
  \bibfield  {author} {\bibinfo {author} {\bibfnamefont {L.}~\bibnamefont
  {Pauling}},\ }\bibfield  {title} {\enquote {\bibinfo {title} {The structure
  and entropy of ice and of other crystals with some randomness of atomic
  arrangement},}\ }\href@noop {} {\bibfield  {journal} {\bibinfo  {journal}
  {Journal of the American Chemical Society}\ }\textbf {\bibinfo {volume}
  {57}},\ \bibinfo {pages} {2680--2684} (\bibinfo {year} {1935})}\BibitemShut
  {NoStop}%
\bibitem [{\citenamefont {Tajima}, \citenamefont {Matsuo},\ and\ \citenamefont
  {Suga}(1982)}]{Tajima82}%
  \BibitemOpen
  \bibfield  {author} {\bibinfo {author} {\bibfnamefont {Y.}~\bibnamefont
  {Tajima}}, \bibinfo {author} {\bibfnamefont {T.}~\bibnamefont {Matsuo}}, \
  and\ \bibinfo {author} {\bibfnamefont {H.}~\bibnamefont {Suga}},\ }\bibfield
  {title} {\enquote {\bibinfo {title} {Phase transition in koh-doped hexagonal
  ice},}\ }\href@noop {} {\bibfield  {journal} {\bibinfo  {journal} {Nature}\
  }\textbf {\bibinfo {volume} {299}},\ \bibinfo {pages} {810--812} (\bibinfo
  {year} {1982})}\BibitemShut {NoStop}%
\bibitem [{\citenamefont {Bart{\'o}k}, \citenamefont {Kondor},\ and\
  \citenamefont {Cs{\'a}nyi}(2013)}]{Bartok13}%
  \BibitemOpen
  \bibfield  {author} {\bibinfo {author} {\bibfnamefont {A.~P.}\ \bibnamefont
  {Bart{\'o}k}}, \bibinfo {author} {\bibfnamefont {R.}~\bibnamefont {Kondor}},
  \ and\ \bibinfo {author} {\bibfnamefont {G.}~\bibnamefont {Cs{\'a}nyi}},\
  }\bibfield  {title} {\enquote {\bibinfo {title} {On representing chemical
  environments},}\ }\href@noop {} {\bibfield  {journal} {\bibinfo  {journal}
  {Physical Review B}\ }\textbf {\bibinfo {volume} {87}},\ \bibinfo {pages}
  {184115} (\bibinfo {year} {2013})}\BibitemShut {NoStop}%
\bibitem [{\citenamefont {Bonomi}\ and\ \citenamefont
  {Parrinello}(2010)}]{Bonomi10}%
  \BibitemOpen
  \bibfield  {author} {\bibinfo {author} {\bibfnamefont {M.}~\bibnamefont
  {Bonomi}}\ and\ \bibinfo {author} {\bibfnamefont {M.}~\bibnamefont
  {Parrinello}},\ }\bibfield  {title} {\enquote {\bibinfo {title} {Enhanced
  sampling in the well-tempered ensemble},}\ }\href {\doibase
  10.1103/PhysRevLett.104.190601} {\bibfield  {journal} {\bibinfo  {journal}
  {Physical review letters}\ }\textbf {\bibinfo {volume} {104}},\ \bibinfo
  {pages} {190601} (\bibinfo {year} {2010})}\BibitemShut {NoStop}%
\bibitem [{\citenamefont {Barducci}, \citenamefont {Bussi},\ and\ \citenamefont
  {Parrinello}(2008)}]{Barducci08}%
  \BibitemOpen
  \bibfield  {author} {\bibinfo {author} {\bibfnamefont {A.}~\bibnamefont
  {Barducci}}, \bibinfo {author} {\bibfnamefont {G.}~\bibnamefont {Bussi}}, \
  and\ \bibinfo {author} {\bibfnamefont {M.}~\bibnamefont {Parrinello}},\
  }\bibfield  {title} {\enquote {\bibinfo {title} {Well-tempered metadynamics:
  A smoothly converging and tunable free-energy method},}\ }\href {\doibase
  10.1103/PhysRevLett.100.020603} {\bibfield  {journal} {\bibinfo  {journal}
  {Physical review letters}\ }\textbf {\bibinfo {volume} {100}},\ \bibinfo
  {pages} {020603} (\bibinfo {year} {2008})}\BibitemShut {NoStop}%
\bibitem [{\citenamefont {Newman}\ and\ \citenamefont
  {Barkema}(1999)}]{NewmanBarkemaBook}%
  \BibitemOpen
  \bibfield  {author} {\bibinfo {author} {\bibfnamefont {M.}~\bibnamefont
  {Newman}}\ and\ \bibinfo {author} {\bibfnamefont {G.}~\bibnamefont
  {Barkema}},\ }\href@noop {} {\emph {\bibinfo {title} {Monte carlo methods in
  statistical physics chapter 1-4}}}\ (\bibinfo  {publisher} {Oxford University
  Press: New York, USA},\ \bibinfo {year} {1999})\BibitemShut {NoStop}%
\bibitem [{\citenamefont {Niu}\ \emph {et~al.}(2018)\citenamefont {Niu},
  \citenamefont {Piaggi}, \citenamefont {Invernizzi},\ and\ \citenamefont
  {Parrinello}}]{Niu18}%
  \BibitemOpen
  \bibfield  {author} {\bibinfo {author} {\bibfnamefont {H.}~\bibnamefont
  {Niu}}, \bibinfo {author} {\bibfnamefont {P.~M.}\ \bibnamefont {Piaggi}},
  \bibinfo {author} {\bibfnamefont {M.}~\bibnamefont {Invernizzi}}, \ and\
  \bibinfo {author} {\bibfnamefont {M.}~\bibnamefont {Parrinello}},\ }\bibfield
   {title} {\enquote {\bibinfo {title} {Molecular dynamics simulations of
  liquid silica crystallization},}\ }\href@noop {} {\bibfield  {journal}
  {\bibinfo  {journal} {Proceedings of the National Academy of Sciences}\
  }\textbf {\bibinfo {volume} {115}},\ \bibinfo {pages} {5348--5352} (\bibinfo
  {year} {2018})}\BibitemShut {NoStop}%
\bibitem [{\citenamefont {Hansmann}(1997)}]{Hansmann97}%
  \BibitemOpen
  \bibfield  {author} {\bibinfo {author} {\bibfnamefont {U.~H.}\ \bibnamefont
  {Hansmann}},\ }\bibfield  {title} {\enquote {\bibinfo {title} {Parallel
  tempering algorithm for conformational studies of biological molecules},}\
  }\href@noop {} {\bibfield  {journal} {\bibinfo  {journal} {Chemical Physics
  Letters}\ }\textbf {\bibinfo {volume} {281}},\ \bibinfo {pages} {140--150}
  (\bibinfo {year} {1997})}\BibitemShut {NoStop}%
\bibitem [{\citenamefont {Sugita}\ and\ \citenamefont
  {Okamoto}(1999)}]{Sugita99}%
  \BibitemOpen
  \bibfield  {author} {\bibinfo {author} {\bibfnamefont {Y.}~\bibnamefont
  {Sugita}}\ and\ \bibinfo {author} {\bibfnamefont {Y.}~\bibnamefont
  {Okamoto}},\ }\bibfield  {title} {\enquote {\bibinfo {title}
  {Replica-exchange molecular dynamics method for protein folding},}\
  }\href@noop {} {\bibfield  {journal} {\bibinfo  {journal} {Chemical physics
  letters}\ }\textbf {\bibinfo {volume} {314}},\ \bibinfo {pages} {141--151}
  (\bibinfo {year} {1999})}\BibitemShut {NoStop}%
\bibitem [{\citenamefont {Yang}, \citenamefont {Niu},\ and\ \citenamefont
  {Parrinello}(2018)}]{Yang18}%
  \BibitemOpen
  \bibfield  {author} {\bibinfo {author} {\bibfnamefont {Y.~I.}\ \bibnamefont
  {Yang}}, \bibinfo {author} {\bibfnamefont {H.}~\bibnamefont {Niu}}, \ and\
  \bibinfo {author} {\bibfnamefont {M.}~\bibnamefont {Parrinello}},\ }\bibfield
   {title} {\enquote {\bibinfo {title} {Combining metadynamics and integrated
  tempering sampling},}\ }\href {\doibase 10.1021/acs.jpclett.8b03005}
  {\bibfield  {journal} {\bibinfo  {journal} {The Journal of Physical Chemistry
  Letters}\ }\textbf {\bibinfo {volume} {9}},\ \bibinfo {pages} {6426}
  (\bibinfo {year} {2018})}\BibitemShut {NoStop}%
\bibitem [{\citenamefont {Binder}(1987)}]{Binder87}%
  \BibitemOpen
  \bibfield  {author} {\bibinfo {author} {\bibfnamefont {K.}~\bibnamefont
  {Binder}},\ }\bibfield  {title} {\enquote {\bibinfo {title} {Theory of
  first-order phase transitions},}\ }\href@noop {} {\bibfield  {journal}
  {\bibinfo  {journal} {Reports on progress in physics}\ }\textbf {\bibinfo
  {volume} {50}},\ \bibinfo {pages} {783} (\bibinfo {year} {1987})}\BibitemShut
  {NoStop}%
\bibitem [{\citenamefont {Vega}, \citenamefont {Martin-Conde},\ and\
  \citenamefont {Patrykiejew}(2006)}]{Vega06}%
  \BibitemOpen
  \bibfield  {author} {\bibinfo {author} {\bibfnamefont {C.}~\bibnamefont
  {Vega}}, \bibinfo {author} {\bibfnamefont {M.}~\bibnamefont {Martin-Conde}},
  \ and\ \bibinfo {author} {\bibfnamefont {A.}~\bibnamefont {Patrykiejew}},\
  }\bibfield  {title} {\enquote {\bibinfo {title} {Absence of superheating for
  ice ih with a free surface: A new method of determining the melting point of
  different water models},}\ }\href@noop {} {\bibfield  {journal} {\bibinfo
  {journal} {Molecular Physics}\ }\textbf {\bibinfo {volume} {104}},\ \bibinfo
  {pages} {3583--3592} (\bibinfo {year} {2006})}\BibitemShut {NoStop}%
\bibitem [{\citenamefont {Rahman}\ and\ \citenamefont
  {Stillinger}(1972)}]{Rahman72}%
  \BibitemOpen
  \bibfield  {author} {\bibinfo {author} {\bibfnamefont {A.}~\bibnamefont
  {Rahman}}\ and\ \bibinfo {author} {\bibfnamefont {F.~H.}\ \bibnamefont
  {Stillinger}},\ }\bibfield  {title} {\enquote {\bibinfo {title} {Proton
  distribution in ice and the kirkwood correlation factor},}\ }\href@noop {}
  {\bibfield  {journal} {\bibinfo  {journal} {The Journal of Chemical Physics}\
  }\textbf {\bibinfo {volume} {57}},\ \bibinfo {pages} {4009--4017} (\bibinfo
  {year} {1972})}\BibitemShut {NoStop}%
\bibitem [{\citenamefont {Plimpton}(1995)}]{Plimpton95}%
  \BibitemOpen
  \bibfield  {author} {\bibinfo {author} {\bibfnamefont {S.}~\bibnamefont
  {Plimpton}},\ }\bibfield  {title} {\enquote {\bibinfo {title} {Fast parallel
  algorithms for short-range molecular dynamics},}\ }\href {\doibase
  10.1006/jcph.1995.1039} {\bibfield  {journal} {\bibinfo  {journal} {Journal
  of computational physics}\ }\textbf {\bibinfo {volume} {117}},\ \bibinfo
  {pages} {1--19} (\bibinfo {year} {1995})}\BibitemShut {NoStop}%
\bibitem [{\citenamefont {Tribello}\ \emph {et~al.}(2014)\citenamefont
  {Tribello}, \citenamefont {Bonomi}, \citenamefont {Branduardi}, \citenamefont
  {Camilloni},\ and\ \citenamefont {Bussi}}]{Tribello14}%
  \BibitemOpen
  \bibfield  {author} {\bibinfo {author} {\bibfnamefont {G.~A.}\ \bibnamefont
  {Tribello}}, \bibinfo {author} {\bibfnamefont {M.}~\bibnamefont {Bonomi}},
  \bibinfo {author} {\bibfnamefont {D.}~\bibnamefont {Branduardi}}, \bibinfo
  {author} {\bibfnamefont {C.}~\bibnamefont {Camilloni}}, \ and\ \bibinfo
  {author} {\bibfnamefont {G.}~\bibnamefont {Bussi}},\ }\bibfield  {title}
  {\enquote {\bibinfo {title} {Plumed 2: New feathers for an old bird},}\
  }\href {\doibase 10.1016/j.cpc.2013.09.018} {\bibfield  {journal} {\bibinfo
  {journal} {Computer Physics Communications}\ }\textbf {\bibinfo {volume}
  {185}},\ \bibinfo {pages} {604--613} (\bibinfo {year} {2014})}\BibitemShut
  {NoStop}%
\bibitem [{ves()}]{vescode}%
  \BibitemOpen
  \href {http://www.ves-code.org} {}\bibinfo {note} {\textit{VES Code}, a
  library that implements enhanced sampling methods based on Variationally
  Enhanced Sampling written by O.\ Valsson. For the current version, see
  http://www.ves-code.org}\BibitemShut {NoStop}%
\bibitem [{\citenamefont {Bonomi}\ \emph {et~al.}(2019)\citenamefont {Bonomi},
  \citenamefont {Bussi}, \citenamefont {Camilloni}, \citenamefont {Tribello},
  \citenamefont {Ban{\'a}{\v{s}}}, \citenamefont {Barducci}, \citenamefont
  {Bernetti}, \citenamefont {Bolhuis}, \citenamefont {Bottaro}, \citenamefont
  {Branduardi} \emph {et~al.}}]{Bonomi19}%
  \BibitemOpen
  \bibfield  {author} {\bibinfo {author} {\bibfnamefont {M.}~\bibnamefont
  {Bonomi}}, \bibinfo {author} {\bibfnamefont {G.}~\bibnamefont {Bussi}},
  \bibinfo {author} {\bibfnamefont {C.}~\bibnamefont {Camilloni}}, \bibinfo
  {author} {\bibfnamefont {G.~A.}\ \bibnamefont {Tribello}}, \bibinfo {author}
  {\bibfnamefont {P.}~\bibnamefont {Ban{\'a}{\v{s}}}}, \bibinfo {author}
  {\bibfnamefont {A.}~\bibnamefont {Barducci}}, \bibinfo {author}
  {\bibfnamefont {M.}~\bibnamefont {Bernetti}}, \bibinfo {author}
  {\bibfnamefont {P.~G.}\ \bibnamefont {Bolhuis}}, \bibinfo {author}
  {\bibfnamefont {S.}~\bibnamefont {Bottaro}}, \bibinfo {author} {\bibfnamefont
  {D.}~\bibnamefont {Branduardi}},  \emph {et~al.},\ }\bibfield  {title}
  {\enquote {\bibinfo {title} {Promoting transparency and reproducibility in
  enhanced molecular simulations},}\ }\href@noop {} {\bibfield  {journal}
  {\bibinfo  {journal} {Nature methods}\ }\textbf {\bibinfo {volume} {16}},\
  \bibinfo {pages} {670--673} (\bibinfo {year} {2019})}\BibitemShut {NoStop}%
\bibitem [{\citenamefont {Bussi}, \citenamefont {Donadio},\ and\ \citenamefont
  {Parrinello}(2007)}]{Bussi07}%
  \BibitemOpen
  \bibfield  {author} {\bibinfo {author} {\bibfnamefont {G.}~\bibnamefont
  {Bussi}}, \bibinfo {author} {\bibfnamefont {D.}~\bibnamefont {Donadio}}, \
  and\ \bibinfo {author} {\bibfnamefont {M.}~\bibnamefont {Parrinello}},\
  }\bibfield  {title} {\enquote {\bibinfo {title} {Canonical sampling through
  velocity rescaling},}\ }\href {\doibase 10.1063/1.2408420} {\bibfield
  {journal} {\bibinfo  {journal} {The Journal of chemical physics}\ }\textbf
  {\bibinfo {volume} {126}},\ \bibinfo {pages} {014101} (\bibinfo {year}
  {2007})}\BibitemShut {NoStop}%
\bibitem [{\citenamefont {Parrinello}\ and\ \citenamefont
  {Rahman}(1981)}]{Parrinello81}%
  \BibitemOpen
  \bibfield  {author} {\bibinfo {author} {\bibfnamefont {M.}~\bibnamefont
  {Parrinello}}\ and\ \bibinfo {author} {\bibfnamefont {A.}~\bibnamefont
  {Rahman}},\ }\bibfield  {title} {\enquote {\bibinfo {title} {Polymorphic
  transitions in single crystals: A new molecular dynamics method},}\ }\href
  {\doibase 10.1063/1.328693} {\bibfield  {journal} {\bibinfo  {journal}
  {Journal of Applied physics}\ }\textbf {\bibinfo {volume} {52}},\ \bibinfo
  {pages} {7182--7190} (\bibinfo {year} {1981})}\BibitemShut {NoStop}%
\bibitem [{\citenamefont {Ryckaert}, \citenamefont {Ciccotti},\ and\
  \citenamefont {Berendsen}(1977)}]{Ryckaert77}%
  \BibitemOpen
  \bibfield  {author} {\bibinfo {author} {\bibfnamefont {J.-P.}\ \bibnamefont
  {Ryckaert}}, \bibinfo {author} {\bibfnamefont {G.}~\bibnamefont {Ciccotti}},
  \ and\ \bibinfo {author} {\bibfnamefont {H.~J.}\ \bibnamefont {Berendsen}},\
  }\bibfield  {title} {\enquote {\bibinfo {title} {Numerical integration of the
  cartesian equations of motion of a system with constraints: molecular
  dynamics of n-alkanes},}\ }\href@noop {} {\bibfield  {journal} {\bibinfo
  {journal} {Journal of computational physics}\ }\textbf {\bibinfo {volume}
  {23}},\ \bibinfo {pages} {327--341} (\bibinfo {year} {1977})}\BibitemShut
  {NoStop}%
\bibitem [{\citenamefont {Hockney}\ and\ \citenamefont
  {Eastwood}(1988)}]{Hockney88}%
  \BibitemOpen
  \bibfield  {author} {\bibinfo {author} {\bibfnamefont {R.~W.}\ \bibnamefont
  {Hockney}}\ and\ \bibinfo {author} {\bibfnamefont {J.~W.}\ \bibnamefont
  {Eastwood}},\ }\href@noop {} {\emph {\bibinfo {title} {Computer simulation
  using particles}}}\ (\bibinfo  {publisher} {crc Press},\ \bibinfo {year}
  {1988})\BibitemShut {NoStop}%
\bibitem [{\citenamefont {Frenkel}\ and\ \citenamefont
  {Smit}(2001)}]{FrenkelBook}%
  \BibitemOpen
  \bibfield  {author} {\bibinfo {author} {\bibfnamefont {D.}~\bibnamefont
  {Frenkel}}\ and\ \bibinfo {author} {\bibfnamefont {B.}~\bibnamefont {Smit}},\
  }\href@noop {} {\emph {\bibinfo {title} {Understanding molecular simulation:
  from algorithms to applications}}},\ Vol.~\bibinfo {volume} {1}\ (\bibinfo
  {publisher} {Academic press},\ \bibinfo {year} {2001})\BibitemShut {NoStop}%
\bibitem [{\citenamefont {Valsson}\ and\ \citenamefont
  {Parrinello}(2015)}]{Valsson15}%
  \BibitemOpen
  \bibfield  {author} {\bibinfo {author} {\bibfnamefont {O.}~\bibnamefont
  {Valsson}}\ and\ \bibinfo {author} {\bibfnamefont {M.}~\bibnamefont
  {Parrinello}},\ }\bibfield  {title} {\enquote {\bibinfo {title}
  {Well-tempered variational approach to enhanced sampling},}\ }\href {\doibase
  10.1021/acs.jctc.5b00076} {\bibfield  {journal} {\bibinfo  {journal} {Journal
  of chemical theory and computation}\ }\textbf {\bibinfo {volume} {11}},\
  \bibinfo {pages} {1996--2002} (\bibinfo {year} {2015})}\BibitemShut {NoStop}%
\bibitem [{\citenamefont {Bach}\ and\ \citenamefont {Moulines}(2013)}]{Bach13}%
  \BibitemOpen
  \bibfield  {author} {\bibinfo {author} {\bibfnamefont {F.}~\bibnamefont
  {Bach}}\ and\ \bibinfo {author} {\bibfnamefont {E.}~\bibnamefont
  {Moulines}},\ }\bibfield  {title} {\enquote {\bibinfo {title}
  {Non-strongly-convex smooth stochastic approximation with convergence rate o
  (1/n)},}\ }in\ \href@noop {} {\emph {\bibinfo {booktitle} {Advances in Neural
  Information Processing Systems}}}\ (\bibinfo {year} {2013})\ pp.\ \bibinfo
  {pages} {773--781}\BibitemShut {NoStop}%
\bibitem [{\citenamefont {Raiteri}\ \emph {et~al.}(2006)\citenamefont
  {Raiteri}, \citenamefont {Laio}, \citenamefont {Gervasio}, \citenamefont
  {Micheletti},\ and\ \citenamefont {Parrinello}}]{Raiteri06}%
  \BibitemOpen
  \bibfield  {author} {\bibinfo {author} {\bibfnamefont {P.}~\bibnamefont
  {Raiteri}}, \bibinfo {author} {\bibfnamefont {A.}~\bibnamefont {Laio}},
  \bibinfo {author} {\bibfnamefont {F.~L.}\ \bibnamefont {Gervasio}}, \bibinfo
  {author} {\bibfnamefont {C.}~\bibnamefont {Micheletti}}, \ and\ \bibinfo
  {author} {\bibfnamefont {M.}~\bibnamefont {Parrinello}},\ }\bibfield  {title}
  {\enquote {\bibinfo {title} {Efficient reconstruction of complex free energy
  landscapes by multiple walkers metadynamics},}\ }\href@noop {} {\bibfield
  {journal} {\bibinfo  {journal} {The Journal of Physical Chemistry B}\
  }\textbf {\bibinfo {volume} {110}},\ \bibinfo {pages} {3533--3539} (\bibinfo
  {year} {2006})}\BibitemShut {NoStop}%
\bibitem [{\citenamefont {Torrie}\ and\ \citenamefont
  {Valleau}(1977)}]{Torrie77}%
  \BibitemOpen
  \bibfield  {author} {\bibinfo {author} {\bibfnamefont {G.~M.}\ \bibnamefont
  {Torrie}}\ and\ \bibinfo {author} {\bibfnamefont {J.~P.}\ \bibnamefont
  {Valleau}},\ }\bibfield  {title} {\enquote {\bibinfo {title} {Nonphysical
  sampling distributions in monte carlo free-energy estimation: Umbrella
  sampling},}\ }\href@noop {} {\bibfield  {journal} {\bibinfo  {journal}
  {Journal of Computational Physics}\ }\textbf {\bibinfo {volume} {23}},\
  \bibinfo {pages} {187--199} (\bibinfo {year} {1977})}\BibitemShut {NoStop}%
\bibitem [{\citenamefont {Tribello}\ \emph {et~al.}(2017)\citenamefont
  {Tribello}, \citenamefont {Giberti}, \citenamefont {Sosso}, \citenamefont
  {Salvalaglio},\ and\ \citenamefont {Parrinello}}]{Tribello17}%
  \BibitemOpen
  \bibfield  {author} {\bibinfo {author} {\bibfnamefont {G.~A.}\ \bibnamefont
  {Tribello}}, \bibinfo {author} {\bibfnamefont {F.}~\bibnamefont {Giberti}},
  \bibinfo {author} {\bibfnamefont {G.~C.}\ \bibnamefont {Sosso}}, \bibinfo
  {author} {\bibfnamefont {M.}~\bibnamefont {Salvalaglio}}, \ and\ \bibinfo
  {author} {\bibfnamefont {M.}~\bibnamefont {Parrinello}},\ }\bibfield  {title}
  {\enquote {\bibinfo {title} {Analyzing and driving cluster formation in
  atomistic simulations},}\ }\href@noop {} {\bibfield  {journal} {\bibinfo
  {journal} {Journal of chemical theory and computation}\ }\textbf {\bibinfo
  {volume} {13}},\ \bibinfo {pages} {1317--1327} (\bibinfo {year}
  {2017})}\BibitemShut {NoStop}%
\bibitem [{\citenamefont {Steinhardt}, \citenamefont {Nelson},\ and\
  \citenamefont {Ronchetti}(1983)}]{Steinhardt83}%
  \BibitemOpen
  \bibfield  {author} {\bibinfo {author} {\bibfnamefont {P.}~\bibnamefont
  {Steinhardt}}, \bibinfo {author} {\bibfnamefont {D.}~\bibnamefont {Nelson}},
  \ and\ \bibinfo {author} {\bibfnamefont {M.}~\bibnamefont {Ronchetti}},\
  }\bibfield  {title} {\enquote {\bibinfo {title} {Bond-orientational order in
  liquids and glasses},}\ }\href {\doibase 10.1103/PhysRevB.28.784} {\bibfield
  {journal} {\bibinfo  {journal} {Phys. Rev. B}\ }\textbf {\bibinfo {volume}
  {28}},\ \bibinfo {pages} {784--805} (\bibinfo {year} {1983})}\BibitemShut
  {NoStop}%
\bibitem [{\citenamefont {Van~Duijneveldt}\ and\ \citenamefont
  {Frenkel}(1992)}]{vanDuijneveldt92}%
  \BibitemOpen
  \bibfield  {author} {\bibinfo {author} {\bibfnamefont {J.}~\bibnamefont
  {Van~Duijneveldt}}\ and\ \bibinfo {author} {\bibfnamefont {D.}~\bibnamefont
  {Frenkel}},\ }\bibfield  {title} {\enquote {\bibinfo {title} {Computer
  simulation study of free energy barriers in crystal nucleation},}\
  }\href@noop {} {\bibfield  {journal} {\bibinfo  {journal} {The Journal of
  chemical physics}\ }\textbf {\bibinfo {volume} {96}},\ \bibinfo {pages}
  {4655--4668} (\bibinfo {year} {1992})}\BibitemShut {NoStop}%
\bibitem [{\citenamefont {Piaggi}(2020)}]{GitHubRepo}%
  \BibitemOpen
  \bibfield  {author} {\bibinfo {author} {\bibfnamefont {P.}~\bibnamefont
  {Piaggi}},\ }\href {https://github.com/PabloPiaggi/Crystallization-of-IceIh}
  {\enquote {\bibinfo {title}
  {https://github.com/pablopiaggi/crystallization-of-iceih},}\ } (\bibinfo
  {year} {2020})\BibitemShut {NoStop}%
\end{thebibliography}
\end{document}